# The role of nonlinear absorption of an ECR beam for fusion plasma pre-ionization

Tulchhi Ram[1,2], T. Wauters[3], P.C. de Vries[3], P.K. Sharma[1,2], Raju Daniel[1,2]

[1] **Institute for Plasma Research, Bhat, Gandhinagar-382428, India**
[2] **Homi Bhabha National Institute, Anushaktinagar, Mumbai-400094, India**
[3] **ITER Organization, Route de Vinon-Sur-Verdon, CS 90 046, 13067 St Paul Lez Durance Cedex, France**

E-mail: tulchhi.ram@ipr.res.in

**Abstract**

This study investigates the nonlinear interactions between electrons and electron-cyclotron resonance (ECR) heating beams in magnetized, low-temperature fusion plasmas, focusing on enhancing pre-ionization efficiency. We analyse key parameters affecting ECR absorption, including phase angle, temperature inhomogeneity, magnetic field gradients, and beam characteristics like width and frequency. Using 2D simulations for these low temperature plasmas, we demonstrate electrons gain energy through nonlinear trapping in velocity space near the resonance layer, achieving energy levels up to 1 keV under optimal conditions. Notably, narrow beam widths allow electrons to reach higher energy levels more efficiently than broader beams, highlighting the spatial localization of the nonlinear interactions to regions where the field is strong and frequency mismatch is small. Our findings show that in devices with lower-frequency ECR systems, such as TCV (82.7 GHz), electrons can gain up to 800 eV, while higher-frequency devices, like ASDEX-Upgrade (140 GHz) and ITER (170 GHz), achieve lower energy gains (480 eV and 420 eV, respectively) under similar conditions. These insights are directly applicable to improving ECR heating strategies, informing design choices in future fusion devices to optimize energy transfer during plasma breakdown and startup.

## 1. Introduction

In modern magnetic confinement fusion devices, Electron Cyclotron Resonance (ECR) is a crucial auxiliary system to heat plasma discharge. Modern gyrotrons can produce power in the range of 1 MW. Multiple gyrotrons can be used to inject higher powers into the device, each launching a narrow electromagnetic beam of electric fields in the range $10^4 - 10^6$ V/m. The wave energy in these beams is absorbed by the free electrons in the gas or plasma, when a harmonic of the wave frequency is resonant with electron gyration frequency in the device magnetic field. In ITER, both fundamental and second harmonic ECR will be applied[1].

ECR can also be used to assist in the initiation process of plasma discharges[2] or to create low temperature plasmas for the purpose of wall conditioning, so-called Electron Cyclotron Wall Conditioning (ECWC) [3], [4]. Especially, the limitations of the available magnetic flux and the use of continuous toroidal vessels with superconducting coils, which result in lower available electric fields, necessitate secondary methods to initiate plasma discharges. Such plasma initiation techniques, and the effect of ECR on pre-ionization and burnthrough, have been studied and applied in many current fusion devices such as KSTAR[5], JT-60U[6], DIII-D[7], [8], and TCV[9].

During the early stages of plasma initiation, including pre-ionization, burnthrough, and ECWC, the plasma is characterised by low densities and temperatures (a few tens of eV). Under such conditions, ECR beam absorption is limited, allowing waves to pass through the plasma multiple times[10]. While linear theories often suffice to describe wave absorption [11], they may not adequately capture the dynamics of low-density, low-temperature plasma, where nonlinear effects become significant. D. Farina et al. [12]. and Tsironis et al. [13]. have discussed the nonlinear interaction of electrons with EC beams. These studies indicate that electrons can gain significant energy from the ECR beam, even at very low temperatures. A better understanding of the ECR beams with low temperature plasmas may allow a more effective use of ECR for plasma initiation and ECWC. When electrons traverse the resonance region, their perpendicular velocity ($v_\perp$) is strongly modulated by the beam's perpendicular electric field component. Cyclotron resonance heating is typically considered a random process due to phase disruptions caused by collisions. However, in collisionless plasmas, the passage of electrons through resonance zones, coupled with the spatial inhomogeneity of microwave beam intensity, leads to complex dynamics that necessitate a nonlinear treatment of energy absorption.

Building on previous studies exploring linear and nonlinear ECR wave absorption, this work investigates the wave particle interaction for both fundamental and second harmonic electron cyclotron resonance in magnetically confined, low-temperature fusion plasmas. By examining the influence of parameters such as phase angle, beam width, frequency mismatch, and magnetic field gradients, this study provides new insights into optimizing pre-ionization efficiency and energy transfer during plasma initiation.

Section 2 details the governing equation of motion for electron dynamics in presence of the ECR beam. Section 3 provides details of the computational approach, including the 3D-3V simulation methodology to capture the cumulative effects of different parameters in collisionless plasma and key parameter variations such as frequency due to toroidal magnetic field gradient and inhomogeneity in the beam electric field in collisionless plasmas. Special attention is given to the parameters relevant to devices like TCV and ITER. Finally, Section 4 summarizes the findings, discusses their implications for improving ECR heating strategies, and outlines future directions for optimizing ECR systems in fusion devices.

## 2. Equation of motion for the electron

The dynamics of electrons under the influence of ECR heating beams are governed by the relativistic equation of motion, which describes their interactions with electric and magnetic fields. In this study, we consider an electromagnetic wave beam propagating radially with a wave vector $\mathbf{k} = \hat{k}\hat{x}$, exhibiting a Gaussian intensity profile in $y - z$ plane. The electric field of the electron cyclotron resonance (ECR) beam is modelled as $E_w(x, z, t)\hat{y}$, oscillating with time and space. Electrons are assumed to move primarily parallel to the toroidal magnetic field lines, which are oriented along the $z-$axis, while simultaneously spiraling around these field lines. As electrons traverse the Gaussian beam of the electromagnetic wave, their motion is influenced by the oscillating electric field, leading to energy absorption and modulation of their trajectories along the helical path.

The motion of an electron in the presence of a magnetic field and an oscillating electric field is governed by the relativistic equation of motion[14]:

$$\frac{dp}{dt} = e[E_w(x, z, t) + v \times B(x, t)] \quad (1)$$

Where p is the kinetic momentum expressed as $p = m\gamma v$, and $\gamma = \sqrt{1 + \frac{p^2}{m_e^2 c^2}}$ is the relativistic factor. The beam electric field is given by $E_w = E_o(x, z, t)\hat{y}$, while the magnetic field is expressed as $B(x, z, t) = B_\phi(x) + B_w(x, t)$ with $B_\phi$ is the toroidal magnetic field in z-axis and $B_w(x, t)$ represents the wave magnetic field. The equation of motion in these specific conditions can be written as a complex force equation with perpendicular and parallel component given by equation (2) and (3):

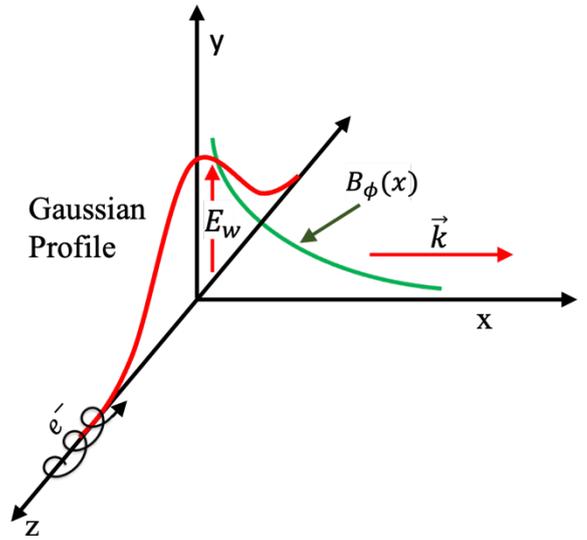

Figure 1 Schematic representation illustrating the Gaussian profile of wave electric field ($E_w$), the gyration motion of an electron along the z-axis, and the wave vector ($\vec{k}$) directed along the x-axis.

$$\frac{dp_\perp}{dt} = ie[E_w(x, z, t) - v_\perp B(x, z, t)] \quad (2)$$

$$\frac{dp_\parallel}{dt} = 0 \quad (3)$$

While the electron's position in the perpendicular plane evolves as equation (4):

$$\frac{d\mathbf{r}}{dt} = \mathbf{v}_\perp \quad (4)$$

The total velocity satisfies $v^2 = v_\perp^2 + v_z^2$, where perpendicular velocity is $v_\perp = v_x + iv_y$, and parallel velocity is $v_z = v_z\hat{z}$. The position in the perpendicular plane is defined as $r = x + iy$. The perpendicular and parallel velocity components are given as $v_\perp = \sqrt{\frac{k_B T_{e,\perp}}{m_e}}$ and $v_\parallel = \sqrt{\frac{k_B T_{e,\parallel}}{m_e}}$ respectively. The position of the electrons along the field lines is described as $z = v_z t$. The toroidal magnetic field magnetic field dependency in $\hat{x}$ direction is expressed as $B_z = \frac{B_o R}{R+X}\hat{z}$ and corresponding electron cyclotron frequency ($\omega_{ce}$) is given by:

$$\omega_{ce}(x) = \frac{\omega_{ce0} R_o}{R_o + X}$$

where $\omega_{ceo} = eB_o/m_e$ with $\omega = \omega_{ceo}$, e is the electron charge, $m_e$ the electron mass, $R_o$ represents the radial location of the resonance, while $X = x_0 + x(t)$ defines the position relative to the resonance with $x_0 = 0$ at the resonance and x(t) oscillatory position of electron in radial direction due to perpendicular motion. The electron cyclotron resonance is possible if $\omega = n\omega_{ceo}$, where $n$ is the harmonic number. The electric field amplitude of the wave, assuming right-handed polarization, is expressed as:

$$E_w = Re\left(E_o e^{-\frac{z^2}{\Delta w^2} + i\chi}\right) \quad (5)$$

Where $\chi = k_\perp x - \omega t + \delta$, with $k_\perp$ the perpendicular wave vector, $k_\parallel$ the parallel wave vector, and $\delta$ is the phase angle. In strong magnetic fields and low temperatures, the electron gyro radius is small compared to the beam width in the $\hat{y}$ direction, allowing the variation in y to be neglected. The electric field configurations are chosen to reflect real experimental conditions, where the field is launched radially and oscillates over time with a Gaussian beam profile in $y - z$ plane. The maximum value of the electric field ($E_o$) in beam[12] is given as $E_o = 2\left[\frac{\eta_o P}{\pi \Delta w^2}\right]^2$, where $\Delta w$ is width of gaussian beam, and the vacuum impedance $\eta = 376.73 \, \Omega$ is impedance of vacuum.

## 3. Numerical Results

The model described in the Equation (1) was implemented for tokamak geometry to simulate the finite beam region. The equation of motion was numerically solved using the fourth order Runge-Kutta (RK4) method. Initial simulations focused on a single electron, with one parameter varied at a time to study its individual impact. Subsequently, the study expanded to the R-Z plane, enabling to understand the cumulative effects of parameters on electrons dynamics. This approach allowed us to systematically isolate the effects of individual variables and then observe their combined influence in a realistic tokamak configuration.

### 3.1. Single electron analysis

Figure 2 represents the electron energy ($W = mc^2(\gamma - 1)/e$) as it transverses a beam of power $P = 1 \, MW$, $f = 170$ GHz, and $w_\parallel = 0.025$ m for initial kinetic energies ($W_{in} = 0.03$ eV, 5 eV and 10 eV). It can be seen that the energy of the electron changes several times in case of a very low initial energy ($W_{in} = 0.03$ eV). In contrast, for $W_{in} = 5$ eV and 10 eV, the interaction time with the wave is significantly shorter, leading to a reduced number of oscillations. The oscillation pattern also differs between cases; for lower initial energy, the nonlinear energy oscillations are more prominent and numerous, whereas for higher initial energies, the oscillations are fewer and less pronounced. Despite these differences, it is clear that strong interactions with the wave can result in significant energy gains in the breakdown and start-up phase.

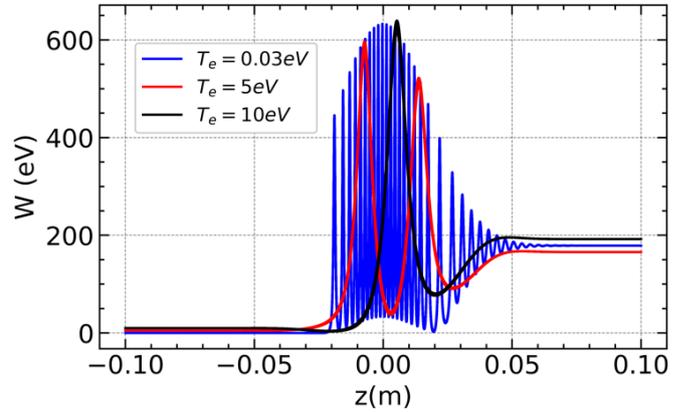

Figure 2 Electron energy for second harmonic as it passes through the beam for $W_{in}$=0.03 eV, 5 eV and 10eV for $f$ =170 GHz, $P = 1$ MW and $w_\parallel = 0.025$ m.

Initially we have explored the phase dependence of the final energy of an electron after a single passing through the EC beam. These parameters will be crucial in understanding the behavior of electron cyclotron resonance (ECR) formed plasmas under these specific conditions. The simulation parameters for the study in Figure 3 are defined as follows: The input power P = 1 MW, with a beam width $w_\parallel = 0.025 \, m$, and the frequency $f$=170 GHz, major radius $R = 6.2$ m, representing an ITER case. Figure 3 shows the variation of the final kinetic energy (W) of electron as a function of the phase angle $\delta$ for $\Delta\omega = \frac{n\omega_{ce}}{\omega} - 1 = 4.3 \times 10^{-3}$ with $n = 2$. The initial kinetic energy is taken to be isotropic with $W_{in} \equiv W_{\parallel,in} = W_{\perp,in}$. The low kinetic energy $w_{in} = 0.03eV$ leads to a final kinetic energy that is more or less independent of the initial phase, $w_{\perp,out} \approx 440$ eV. On increasing the initial kinetic energy of the electron to 5 eV a small and slow variation starts to appear in the final kinetic energy $W$, while over a phase width of $\pi/6$ the final kinetic energy suddenly changes state to very low value. In the slow variation the electron shows around final kinetic energy variation 30 eV variation for $W_{in}$ =5 eV, 60 eV for $W_{in}$ =10 eV, 120 eV for 15 eV, 200 eV for 20 eV and 260 for 25 eV. The final kinetic energy of the electron shows strong variation with $\delta$.

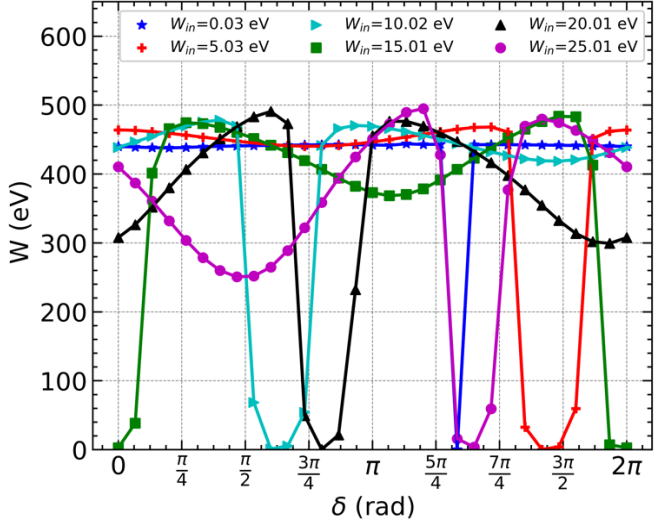

Figure 3 Final kinetic energy (W) of electron passing through the ECR beam and dependence of nonlinear absorption on $\delta$.

Figure 4 shows the variation of average kinetic energy <W (eV) > (averaged over the phase $\delta = 0, 2\pi$) of electrons as a function of the frequency mismatch ($\Delta\omega$) for different beam widths $\Delta w$ and at $W_{in} = 15\ eV$. On increasing the beam width, the maximum average energy gain starts to decrease. For smaller beam width, the interaction peaks at larger frequency mismatch and the range of frequency mismatch where the nonlinear interaction happens increases. The main parameter is the area under the curve in Figure 4, with peak maximum energy gain decreasing almost $1/\Delta w$ on increasing the width of the ECR beam. The most important insight from this plot is that nonlinear effects, such as significant energy gains for the electrons, are predominantly seen when the beam width is narrow which in the case of like TCV[9] ($\Delta w \sim 0.025 - 0.03$ m) and ASDEX-U[15] ($\Delta w \approx 0.015 - 0.02$ m) is possible. In these cases, the electrons are more likely to interact coherently with the high-energy electromagnetic field, leading to greater energy absorption.

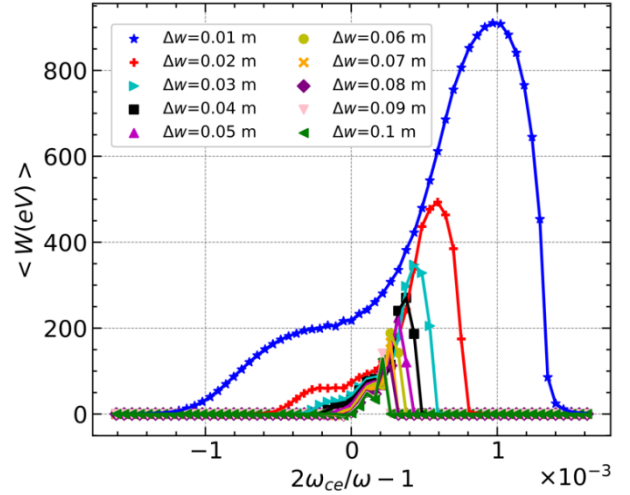

Figure 4 Final kinetic energy (W) of the electron with frequency mismatch $\Delta\omega$ for different beam width ($\Delta w$).

In the context of tokamaks operating at frequencies ranging from 80 to 200 GHz, the simulation results shown in Figure 5 aimed to analyse energy absorption under conditions of frequency mismatch and examine how variations in launched frequency impact energy gain for a fixed power input. To ensure consistent initial conditions, the initial kinetic energies for both parallel and perpendicular directions were set to $W_{\parallel,in} = W_{\perp,in} = 15\ eV$. Figure 5 illustrates the average kinetic energy of electrons over a phase cycle $\boldsymbol{\delta = 0}$ to $\boldsymbol{2\pi}$ offering insights into the behavior of electron energy absorption with frequency mismatch.

The results demonstrate that as the frequency increases, the energy gain tends to diminish due to the growing mismatch between the electron cyclotron resonance (ECR) frequency and the launched frequency. Lower-frequency tokamaks, like TCV at 82.7 GHz, exhibit larger energy gains, with electrons reaching energies up to 780 eV for a frequency mismatch $\boldsymbol{\Delta\omega = 8 \times 10^{-4}}$. This indicates that lower-frequency devices are highly efficient at producing high-energy electrons, for larger frequency mismatch range. Conversely, in higher-frequency devices, such as ASDEX-Upgrade at 140 GHz, the electron energy gains around 480 eV for $\boldsymbol{\Delta\omega = 5.8 \times 10^{-4}}$ showing a reduction in efficiency as the frequency rises.

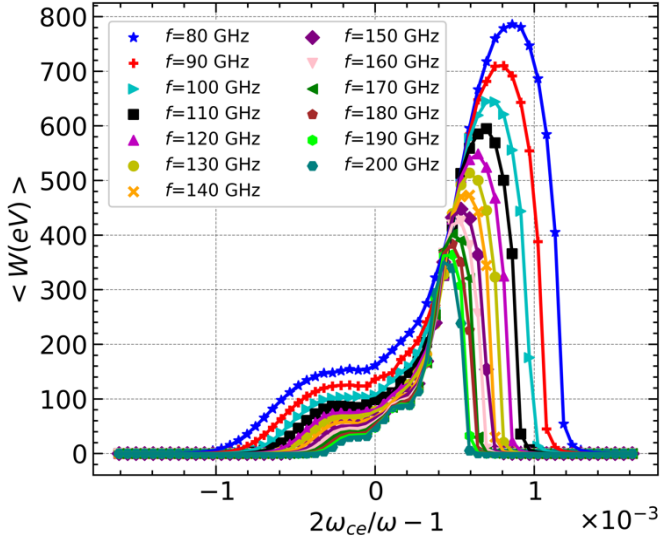

Figure 5 Kinetic energy of electron with frequency mismatch $\frac{2\omega_{ce}}{\omega} - 1$ for different launching frequencies $\Delta w = 0.025$ m, P= 1 MW.

For ITER, which is planned to operates[16], [17] at an even higher frequency of 170 GHz, the maximum energy gain is about 420 eV, lower than both TCV and ASDEX-Upgrade. This further decreases by ITER's larger beam width, ranging from 0.05 to 0.1 m, compared to the narrower beam widths (0.015 to 0.025 m) of TCV and ASDEX-Upgrade. The wider beam leads to lower electric field in the beam for a given power reducing the overall energy gain, with maximum energy gains dropping to around 100 eV for $\Delta w = 0.1$ m. This suggests that ITER, operating at a higher frequency, will produce electrons with lower energy than in present devices, but may produce more energetic electrons due to a larger resonance width as the gradient in the magnetic field is slow compared to the smaller machines.

### 3.2. 2D simulation analysis in x-z plane

The effect of nonlinearity on a large number of electrons has been studied without considering their mutual interactions. The wave's electric field follows the form given by equation (4). A total of 120,000 electrons (electrons) have been used in the simulation. The initial positions of the electrons in the x-z plane are uniformly distributed, while the initial velocities ($v_\perp$, $v_z$) follow a Maxwellian distribution determined by the electron temperature ($T_e$). The simulation domain is set up to clearly show particle interaction with the wave in the z-direction, with a simulation domain width of ($8w_\parallel$) with $\Delta w = 0.05$ m. In the x-direction, the simulation width is chosen to effectively capture the effects within the interaction region. For the first harmonic, where the interaction region is wider, the x-direction width is set to 0.1 m for ITER. For the second harmonic, this width varies with temperature and is kept around $0.015$ m. Within the simulation domain, only the perpendicular particle motion is influenced, leaving the motion in the z-direction unaffected. This causes electrons to reach the boundary in the z-direction and exit the simulation domain, as they continue traveling along z without collisions. To address this, new electrons are introduced near the boundary where electrons are lost, with random velocities assigned according to distribution function. The simulation has been carried out till the energy in the simulation becomes constant or in other words simulation reaches an equilibrium state. As initial electrons from the low temperature start to interact with the wave electric field, the total energy of the system increases but on giving the simulation enough time the energy gain saturates.

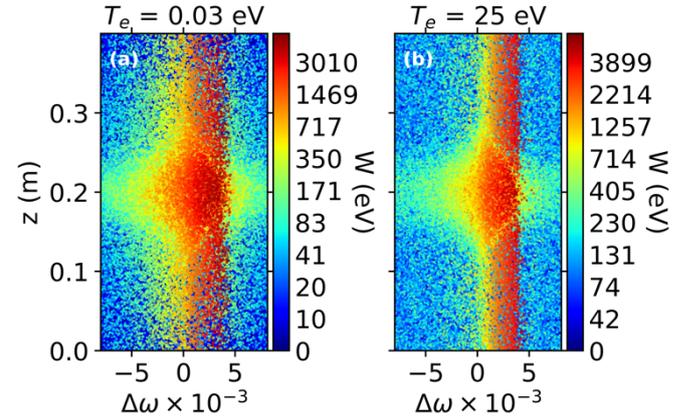

Figure 6 The particle distribution for first harmonic in $x - z$ plane with colours representing particles kinetic energy ($W$) with x axis in terms of $\Delta\omega$ with (a) $T_e = 0.03\ eV$ and (b) $T_e = 25$ eV.

In Figure 6, particle distribution for first harmonic in $x - z$ plane with colours representing particles kinetic energy ($W$) plotted for $f = 170\ GHz$, 1 MW, with $\Delta w = 0.05$ m. The width of the resonance layer is dependent on the electron temperature $T_e$, however, our simulations reveal that for $T_e$= 0.03 eV, the resonance layer becomes wider in the x-direction, suggesting that electrons interact with the wave over a larger frequency mismatch. the frequency mismatch ($\Delta\omega$) for $x = -0.05$ m to $x = 0.05$ m for plotted Figure 6. Interestingly, the maximum energy gains in both temperature cases are nearly equal and align closely with predictions made by D. Farina et al. [12] and given by $W_{\max,n=1} \cong 15.6 \frac{P^{\frac{1}{3}}}{f\Delta w(1-N_\parallel^2)^{2/3}}$. To understand the effect more closely we have plotted the energy of the electrons in the x-direction as shown in Figure 7 (a) $z = 0.38$ m, $T_e = 0.03\ eV$, (b) $z = 0.38$ m, $T_e = 25\ eV$ and (c) $z = 0.2$ m, $T_e = 0.03\ eV$ and (d) $z = 0.2$ m, $T_e = 25\ eV$ with $z = 0.38$ m plane near the boundary of simulation and $z = 0.2\ m$ centre of the peak electric field. Figure 7 (a) and (b) depict the final energy gain where the electrons are not affected by the oscillating electric field, whereas (c) and (d) represent particle behavior within the beam. The electrons kinetic energy is plotted in the z-direction is calculated with $\Delta z = 0.01$ m. The electrons tend to follow a specific energy structure with well-defined energy gains, forming a sharp upper energy gain limit.

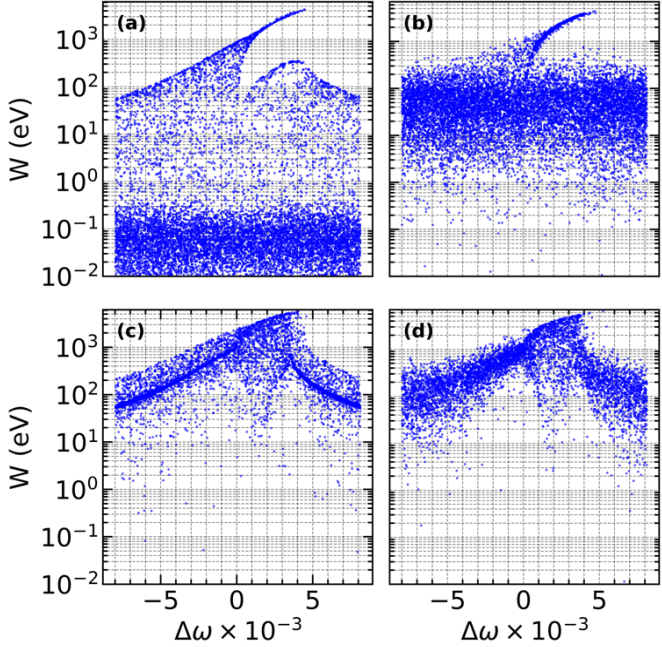

Figure 7 The particle's energy in the radial direction in at (a) $z = 0.38$ m, $T_e = 0.03\ eV$, (b) $z = 0.38$ m, $T_e = 25\ eV$, (c) $z = 0.2$ m, $T_e = 0.03\ eV$ and (d) $z = 0.2$ m, $T_e = 25\ eV$.

For low temperatures, $T_e = 0.03$ eV, we observe that for a given frequency mismatch, two energy limits exist. As shown in Figure 7(a), for $\Delta\omega > 0$, the energy gain exhibits a distinct structure where electrons either end up in a much higher energy state or remain below a lower energy level, which can be up to ten times lower, despite the same magnetic field strength. The upper energy level gained by the electrons is very narrow for higher energies but as we move closer to $\Delta\omega = 0$, the electrons energy gain start to broaden. The electrons gain 100's of eV even for $\Delta\omega > 0$ but a boundary of energy level remains which electrons follow strongly. Nonetheless, even electrons in the lower energy state can reach energies as high as 300 eV and peak of highest energy gained by electrons for lower state and higher energy state is closely coincide near $\Delta\omega \approx 4 \times 10^{-3}$, with the lower energy branch merging with the upper branch near $\Delta\omega = 0$. A similar upper energy branch is found for $T_e = 25$ eV, but the upper branch becomes much narrower, and the tail of the particle near $\Delta\omega = 0$ also become narrow.

Figure 8 shown the analysis of nonlinear collisionless interaction at the second harmonic frequency, as a function of the frequency mismatch with the temperature of up to 30 eV. In Figure 8 (a), (c), (e), (left column of subfigures) has been plotted for $Z = 0.2$ m plane, the centre of the beam. The figure (a) for $T_e = 0.03$ eV shows that the electron interacts with the beam and elevate to higher energy near the $\Delta\omega = -1.5 \times 10^{-4}$ to $1.5 \times 10^{-4}$. Increasing the $T_e = 10\ eV$, figure (c), shows that the elevation in the interaction region reduces and on increasing the temperature up to the 30 eV the number of electrons which are trapped decreases. The electron energy outside the EC beam shows leads at $Z = 0.38$ m (outside the beam) (b), (d), (f) (right column of the subfigures) shows as electrons pass through the beam some electrons gain the energy. On increasing the temperature, the electrons gain higher energies and can interact with the beam for larger frequency mismatch but slowly the structure start to vanish in the background electrons.

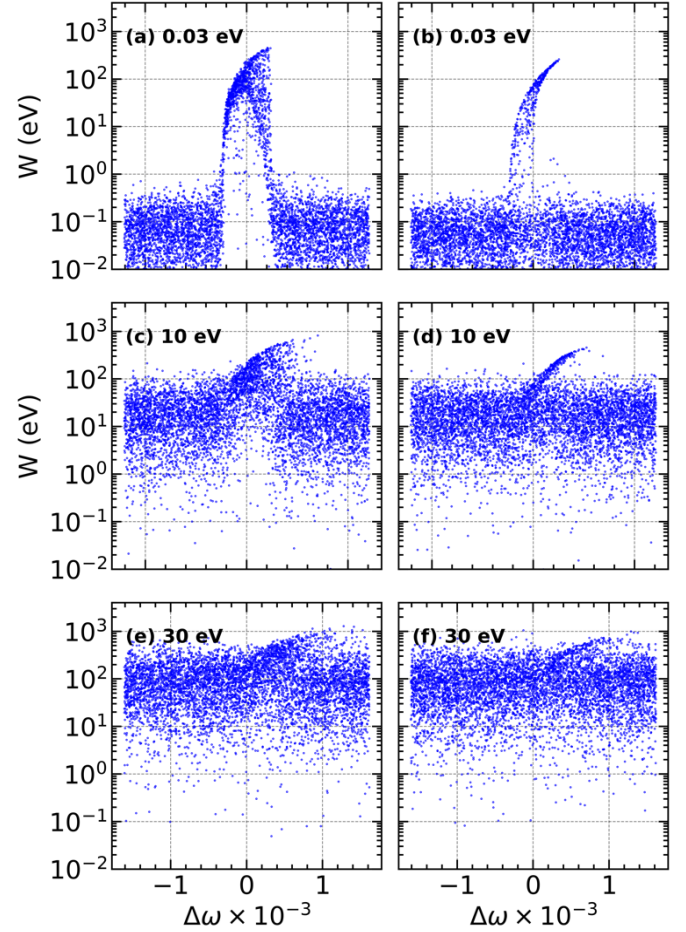

Figure 8 The particle's energy for second harmonic with $\Delta\omega$ for $T_e = 0.03, 10, 30$ eV in $Z = 0.2\ m$ plane $(a, c, e)$ and $Z = 0.38\ m$ plane $(b, d, f)$.

It has been shown that the nonlinear interaction can happen for the first and second harmonic if the parallel velocity ($v_\parallel$) of the particle is for the second harmonic [18]

$$v_\parallel < \frac{w_\parallel}{2\pi}\frac{eE_0}{m_e c} \quad (6)$$

The equation (6) is directly dependent on the wave electric field ($E_o$). The temperature for the second harmonic to this inequality (equation 6) is around 12 eV. Probability of the electrons having kinetic energy below this upper limit for nonlinear interaction can be significant in case of preionization temperatures as shown in Figure 9. Figure 9 shows the fraction of particles with kinetic energy less than 12 eV for a Maxwellian distribution at various electron temperatures. The temperature range used for plotting spans from 0.03 eV to 50 eV. It can be observed that even at an electron temperature of 20 eV, approximately 10% of the particles still have energies below 12 eV. These lower-energy particles may potentially participate in nonlinear interactions.

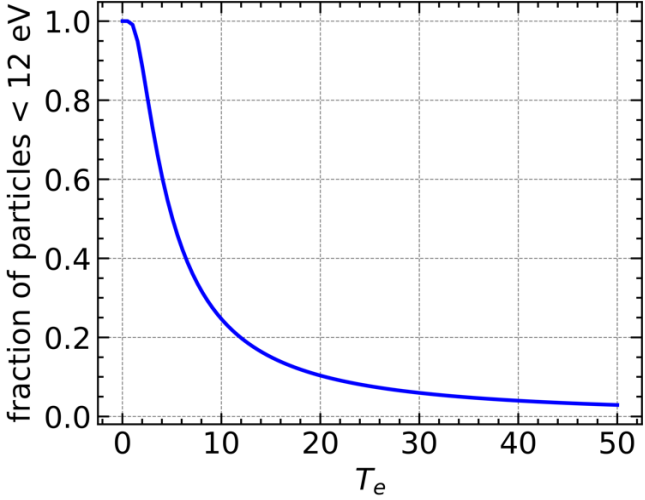

Figure 9 Fraction of particles less then 12 eV with temperature ($T_e$) of the Maxwellian distribution function.

In Figure 10, we observe the energy distribution of electrons in the $x - z$ plane for two different electron temperatures, with an electromagnetic (EM) wave at a frequency of $f = 82.7$ GHz and power of 0.4 MW, and with beam width $w_\parallel = 0.015$ m. In Figure 9 (a), where the electron temperature $T_e = 0.03$ eV, the distribution shows a distinctive rhombus or diamond-like shape, cantered around $z = 0.06$ m. This shape is located within the EM beam, resulting in a concentrated structure with high-energy electrons forming a core region. The energy gain of electrons is highest near $\Delta\omega = 10^{-3}$, indicating resonance-like effects, where electrons absorb maximum energy from the EM field. The perpendicular energy shows strong oscillation trapping in the beam electric field.

In Figure 10 (b), where the electron temperature is increased to $T_e = 25$ eV the particle energy distribution is more spread out and lacks the sharply defined rhombus structure observed in Figure 10 (a). Here, the electrons exhibit a broader energy distribution across both spatial and frequency mismatch domains, showing that at higher temperatures, electrons are more dynamically distributed and less confined by the EM field. This broader distribution indicates that as thermal energy increases, larger velocity leads to fast travel through the beams, and the resonance effect seen at lower temperatures weakens. Consequently, the energy gain is more uniformly spread over a wider range of $\Delta\omega$ and z values, with the trapping effect reduced.

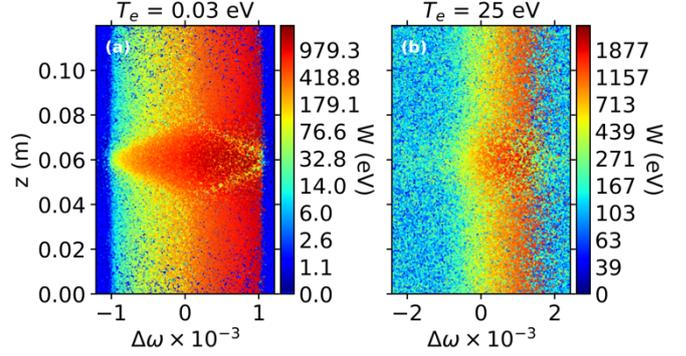

Figure 10 particle distribution for second harmonic in $x - z$ plane with colours representing particles kinetic energy ($W$) with x axis in terms of $\Delta\omega$ with (a) $T_e = 0.03\ eV$ and (b) $T_e = 25$ eV.

In Figure 11 shows the energy of electrons ($W$) with the frequency mismatch ($\Delta\omega$) to understand the effects inside the beam and outside the beam. In Figure 11 (a) and (b), with $z = 0.06$, and $T_e = 0.03$ eV, 25 eV respectively. The $z = 0.06$ m is centre of the beam and to plot these figures we have taken $z = 0.005$ m width in z direction. The Figure 11 (a) shows that the particle inside the intraction zone start to elevate to the higher energy state. Electrons start to oscillate with the beam and remained trapped. In the Figure 11 (b) it can been seen the signature of the trapped region can be seen but the all electrons do not follow the beam electric field. The increased temperature allows wave to interact for larger beam width. Further the same type of the plotting has been done outside the beam region in Figure 11 (c) and (d) at $z = 0.01$ m for $T_e = 0.03$ eV and 25 eV. In Figure 11 (c) electrons shows as specific structure of energy gain. The energy gain for $\Delta\omega < 0$ seen broadening in the energy from few 10's of eV to around 330 eV near the $\Delta\omega = 0$. The energy gain for $\Delta\omega > 0$, the energy starts to concentrate in narrow energy width. The maximum energy of the particle for $\Delta\omega = 10^{-3}$ is around 1.10 keV, which matches with the maximum energy gain $W_{max,n=2} \cong 2.1 \frac{P^{\frac{1}{2}}}{fw_\parallel(1-N_\parallel^2)^{\frac{1}{2}}}$ with $N_\parallel = 0$, the $W_{max,n=2} = 1.07\ keV$. The Figure 11 (d) shows that the structure remains for $T_e = 25$ eV, and maximum energy gain reaches to 2.02 keV near the $\Delta\omega \approx 2.2 \times 10^{-3}$. Unlike the first harmonic the absorption shows strong dependence on the temperature on the final energy gain.

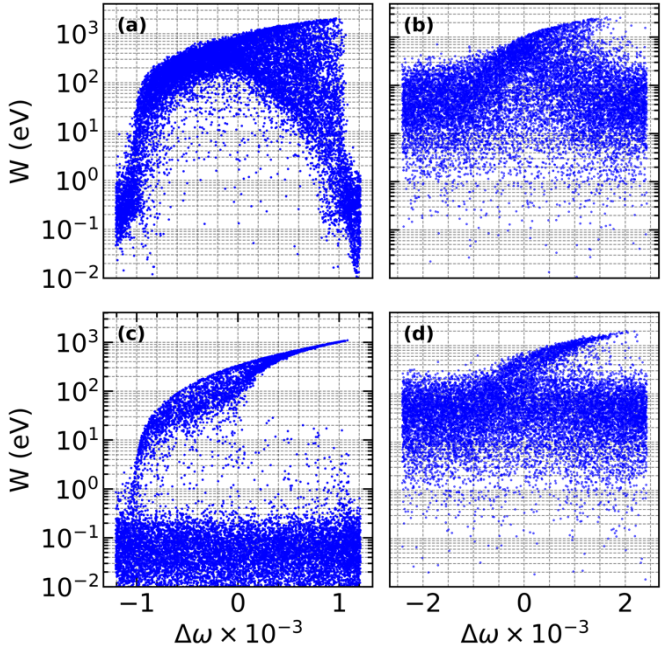

Figure 11 The particle's energy with $\Delta\omega$ at (a) $z = 0.06$ m, $T_e = 0.03\ eV$, (b) $z = 0.06$ m, $T_e = 25\ eV$, (c) $z = 0.01$ m, $T_e = 0.03\ eV$ and (d) $z = 0.01$ m, $T_e = 25\ eV$.

For pre-ionization, the ionization rate depends on both electron energy and density. While electrons inside the beam have higher energy, their spatial confinement limits their contribution to ionization across a broader volume. The larger population of energetic electrons outside the beam, having absorbed energy during transit, is likely more critical for ionization, as they can interact with neutral gas over a wider region.

## 5. Conclusions

This study provides a quantitative analysis of nonlinear electron-cyclotron resonance (ECR) interactions in fusion plasmas, particularly during pre-ionization, highlighting the conditions under which electrons experience significant energy gains. By investigating critical parameters such as phase angle, temperature inhomogeneity, magnetic field gradient, beam width, and launching frequency the study reveals how nonlinear absorption mechanisms optimize energy transfer to electrons, a process crucial for effective plasma ignition.

The single electron method provides valuable insights into the fundamental physics of nonlinear ECR absorption, isolating the effects of individual parameters like phase angle, position of particle which causes the frequency mismatch. The single electron study allows precise control over initial conditions, making it ideal for parametric studies. However, it assumes a single electron's behavior is representative, which overlooks collective effects, statistical variations in initial conditions (e.g., Maxwellian velocity distributions), and the spatial distribution of electrons in a realistic plasma. The test-particle (statistical) approach, as implemented in these 2D simulations, captures these complexities, offering a more comprehensive picture of energy distribution and absorption efficiency across a population of electrons. For future devices like ITER, where pre-ionization efficiency depends on generating a sufficient number of energetic electrons to initiate breakdown, the statistical approach is likely more reliable. It accounts for the cumulative effect of many electrons interacting with the ECR beam under varying conditions, which is critical for predicting macroscopic plasma behavior.

The results emphasize that nonlinear wave-particle interactions in ECR heating can be optimized by tailoring beam width, power, and frequency to the specific fusion device parameters. For instance, in devices like TCV and ASDEX-U, a narrow 0.015–0.025 m beam width at resonance leads to greater energy gains, whereas ITER's broader 0.05–0.1 m beam limits peak absorption to around 100 eV. Additionally, tuning the frequency and controlling phase alignment are essential strategies for improving absorption efficiency, especially in high-frequency devices where frequency mismatches are more detrimental to energy gain.

## Data availability statement

The data that support the findings of this study are available upon reasonable request from the authors.